\documentclass{article}
\usepackage{amssymb,latexsym,amsmath}

\newtheorem{theorem}{Theorem}

\begin{document}

\newcommand{\lap}{\bigtriangleup}
\def\be{\begin{equation}}
\def\ee{\end{equation}}
\def\bea{\begin{eqnarray}}
\def\eea{\end{eqnarray}}
\def\beas{\begin{eqnarray*}}
\def\eeas{\end{eqnarray*}}
\def\n#1{\vert #1 \vert}
\def\nn#1{{\Vert #1 \Vert}}

\newcommand{\R}{\mathbb R}
\newcommand{\N}{\mathbb N}
\def\supp{\mathrm{supp}\,} 
\def\sign{\mathrm{sign}\,}
\def\dist{\mathrm{dist}\,}
\def\dt{\partial_t}
\def\ekin{E_\mathrm{kin}}
\def\epot{E_\mathrm{pot}}

\def\C{{\cal C}}
\def\F{{\cal F}}
\def\G{{\cal G}}
\def\H{{\cal H}}
\def\Hc{{\cal H}_C}
\def\Hr{{{\cal H}_r}}
\def\Rm{{{\cal R}_M}}
\def\r{\rho}
\def\s{\sigma}
\def\e{\epsilon}

\def\prfe{\hspace*{\fill} $\Box$

\smallskip \noindent}

\title{Nonlinear Stability of Newtonian Galaxies and Stars 
from a Mathematical Perspective}

\author{ Gerhard Rein\\
Department of Mathematics\\
University of Bayreuth\\
95440 Bayreuth, Germany}
\date{}
\maketitle

\begin{abstract}
The stability of equilibrium configurations
of galaxies or stars are time honored problems in astrophysics.
We present mathematical results on these problems which have in
recent years been obtained by Yan Guo and the author
in the context 
of the Vlasov-Poisson and the Euler-Poisson model.
Based on a careful analysis of the minimization properties
of conserved quantities---the total energy and so-called Casimir 
functionals---non-linear stability results are obtained for a wide 
class of equilibria.

\end{abstract}

\section{Introduction}
\setcounter{equation}{0}

Under suitable idealizing assumptions the time evolution
of a galaxy can be modeled by the Vlasov-Poisson system
\[ 
\dt f + v \cdot \nabla_x f - \nabla U \cdot \nabla_v f = 0,
\]
\[
\lap U = 4\,\pi\,\r,\ \lim_{\n{x}\to \infty} U(t,x) = 0,
\]
\[ 
\r (t,x) =  \int f (t,x,v)\,dv.
\]
Here $f=f(t,x,v)\geq 0$ denotes the density of the stars in phase space,
$t\in \R,\ x,v \in \R^3$ stand for time, position, and velocity, 
$\r = \r(t,x)$ is the induced spatial mass 
density, and $U=U(t,x)$ is the gravitational potential of the galaxy. 

The problem we wish to address is the non-linear stability of 
stationary solutions of this system. Our
approach will automatically address this question
for the Euler-Poisson system as well,
which describes a self-gravitating fluid ball,
i.e., a barotropic star. The latter model is presented in Section~4.

By definition, a given steady state $f_0$ is stable if
for any neighborhood $N$ of $f_0$ there exists another 
neighborhood $M$ of $f_0$ such that any solution of the system
starting in $M$ will remain in $N$ for
all times. 
This is the usual mathematical definition of
Lyapunov stability.
In the case of an infinite dimensional dynamical system such as the 
Vlasov-Poisson system the choice of the proper concept of ``neighborhood'' 
is a non-trivial part of the stability problem.  
We emphasize that no
approach of the solution to the steady state is asserted---that
would be the concept of asymptotic stability. Since the system as stated
does not include dissipative effects an approach
to a particular steady state is not to be expected in a strict sense;
we do in the present notes not enter into the highly interesting questions
of course graining, phase mixing, etc.
The existence of global-in-time solutions of the system
under consideration at least for initial data close to the steady state
is an integral part of the stability assertion. For the Vlasov-Poisson 
system it was shown in  \cite{Pf} that every smooth, compactly supported initial
datum for $f$ launches a unique global-in-time smooth solution;
a fairly short proof due to J.~Schaeffer is given in  \cite{R1a}.

There is a vast astrophysics literature on the stability question.
However, essentially all investigations that we are aware of proceed
via linearization. This approach suffers from at least two major
difficulties: Firstly, there is no general theory for infinite dimensional
dynamical systems as to how to pass from linearized to non-linear stability.
Secondly, it is well known that if $\lambda$ is an eigenvalue,
i.e., the linearized system has a solution of the form $e^{\lambda t} g(x,v)$,
then the same is true
for $-\lambda$. Hence the optimal case regarding
stability occurs if all such eigenvalues are purely imaginary,
which is precisely the case where stability for the non-linear system
does not follow, not even in finitely many  dimensions.

We will prove non-linear stability
of certain steady states by identifying them as minimizers of a conserved
quantity in terms of which the above neighborhoods $N$ and $M$ are then 
defined. More precisely, for a state $f=f(x,v)\geq 0$ we denote the induced 
spatial mass density and potential by
\[
\r_f (x) :=  \int f (x,v)\,dv,\ U_f(x) := - \int \frac{\r_f (y)}{|x-y|} dy, 
\]
and we introduce the functionals
\bea
\ekin (f)
&:=&
\frac{1}{2} \int\!\!\!\!\int \n{v}^2 f(x,v)\,dv\,dx, \label{ekindef} \\
\epot (f)
&:=&
\frac{1}{2} \int U_f(x)\, \r_f(x)\, dx
= -\frac{1}{8 \pi} \int \n{\nabla U_f(x)}^2\,dx, \label{epotdef}
\eea
the kinetic and the potential energy of the state $f$.
The total energy 
\[
\H := \ekin + \epot
\]
is conserved along solutions of the Vlasov-Poisson system,
but it is indefinite, and it has no critical points, i.e., the
linear part in an expansion about
any state $f_0$ with potential $U_0$ does not vanish:
\[
\H(f) = \H(f_0) + 
\int\!\!\!\!\int \left(\frac{1}{2} |v|^2 + U_0 \right)(f-f_0)  \,dv\,dx
- \frac{1}{8 \pi} \int|\nabla U_f-\nabla U_0 |^2 dx .
\]
However, according to Liouville's theorem the characteristic 
flow corresponding 
to the Vlasov equation preserves phase space volume, and hence
for any reasonable function $\Phi$ the so-called Casimir functional
\[
\C(f) := \int\!\!\!\!\int \Phi(f(x,v))\,dv\,dx
\]
is conserved as well. If we expand the energy-Casimir functional 
\[
\Hc := \H + \C
\]
about an isotropic steady state
\[
f_0(x,v) = \phi(E),\ \ E=E(x,v):= \frac{1}{2} |v|^2 + U_0(x),
\]
we find that
\beas
\H_C(f) 
&=& 
\H_C(f_0) + 
\int\!\!\!\!\int (E + \Phi'(f_0))\,(f-f_0)  \,dv\,dx \\
&& 
{}- \frac{1}{8 \pi} \int|\nabla U_f-\nabla U_0 |^2 dx 
+ \frac{1}{2} \int\!\!\!\!\int \Phi''(f_0) (f-f_0)^2 \,dv\,dx 
+ \ldots .
\eeas
At least formally, we can choose $\Phi$ such that $f_0$ 
is a critical point of $\Hc$,
namely $\Phi' = -\phi^{-1}$, provided $\phi$ is invertible.
The essential problem now is the following: 
In order for the steady state to have finite total mass
the function $\phi$ must vanish above a certain cut-off energy. For $\phi^{-1}$
to exist $\phi$ should thus be decreasing, at least on its support.
But then $\Phi''$ is positive and the quadratic part in the expansion
indefinite. Since one would like to use this quadratic part 
for defining the concept of distance or neighborhood, 
the method seems to fail.
This state of affairs had been observed by various authors,
with the proposed conclusion that the energy-Casimir method
does not work for the stellar dynamics case of the Vlasov-Poisson system,
cf.\ for example  \cite{Kan}.
If the issue is the stability of a plasma, the sign of the
source term in the Poisson equation and hence also the one in front of
the potential energy difference in the expansion above is reversed,
and up to some technicalities stability follows, cf.\  \cite{R1}.  

The approach developed by Yan Guo and the author
to overcome this difficulty is as follows. Starting with a
given function $\Phi$ which defines the Casimir functional
we try to minimize the energy-Casimir functional $\Hc$ under
the constraint that only states with a prescribed total mass  $M>0$
are considered. Under suitable assumptions on $\Phi$ a minimizer does exist
in spite of the fact that the quadratic term in the expansion above 
is indefinite.
One can then show that such a minimizer is a non-linearly stable
steady state.
The exact statements of these results will be given in the next section---the
main assumption is that $\Phi$ is strictly convex which is equivalent
to $f_0(x,v) = \phi(E)$ being strictly decreasing.

The crucial step is to prove the existence of a minimizer.
Here we first construct out of the energy-Casimir functional $\Hc$
a reduced functional  $\Hr$  which is defined on the space of
spatial mass densities $\r$ in such a way that there is a one-to-one
correspondence between the minimizers of the two functionals.
This reduced functional is analyzed in Section~3.
The original motivation for introducing it was purely mathematical:
It is defined on a simpler space, and the troublesome 
part of the original functional is the quadratic and negative definite
potential energy, but the latter
depends on the spatial mass density $\r_f$ and not directly on $f$.
The detour via the reduced 
functional has a beautiful pay-off: The minimizers
of the reduced functional are stable steady states of the Euler-Poisson system
with a macroscopic equation of state corresponding to the
microscopic equation of state $f_0 = \phi(E)$ induced by the 
Casimir functional.
Hence via this reduction procedure we obtain a non-linear proof of
what is often referred to as Antonov's First Law  \cite[p.~305]{BT}:
{\em A spherical stellar system with $f_0 = f_0(E)$
and $df_0/dE < 0$ is stable if the barotropic star with the same
equilibrium density distribution is stable}.
This relation to the Euler-Poisson system, i.e., to the stability of 
gaseous stars, is investigated in Section~4.

In Section~5 we give the main arguments leading to
the existence of a minimizer for the reduced functional.
Mathematically, this is the essential and non-trivial part; it can be 
skipped without compromising the understanding of the rest.

To keep the presentation reasonably simple we restrict ourselves
mostly to spherically symmetric, isotropic steady states. However,
the method has also been applied to non-isotropic steady states, 
to axially symmetric ones,
and to disk-like steady states. Some comments on these
extensions together with other related results as well as open questions
are collected in the last section.

To conclude this introduction we should mention that none of the results
presented here are new, although the way they are presented is new. 
The motivation for these notes is to collect in one place
the main features of our method, 
and to present them in such a way that the readers can
hopefully appreciate the ideas involved. For some details
which are not so relevant for the main argument we refer to
existing papers, but mainly our presentation is aimed to be self-contained.
We include almost no references to the astrophysics literature.
This is really {\em not} done out of disrespect but due to the belief that
our method is essentially the first to address the full {\em non-linear}
stability problem for the systems under consideration. 
Should these notes inspire some comments or criticism from the
astrophysics community we would truly appreciate this.

\noindent
{\bf Acknowledgments:}
These notes are an expanded version of my presentation
at the workshop ``Nonlinear Dynamics in Astronomy and Physics''
in November 2004 at the University of Florida. I truly appreciated
the kind invitation to this inspiring event,
as well as the feedback I received there.
The results reported here originate from my collaboration 
with Y.~Guo, Brown University, whom I would like to thank 
for many stimulating discussions.

\section{Nonlinear stability for the Vlasov-Poisson system---statement 
 of the results}
\setcounter{equation}{0}

We fix a Casimir functional $\C$, i.e., a function $\Phi$
such as
\be \label{poly}
\Phi (f)= f^{1+1/k}, f\geq 0,\ \mbox{with}\ 0<k<3/2,
\ee
more generally: $\Phi \in C^1 ([0,\infty[)$ with $\Phi(0)=0=\Phi'(0)$, and

\smallskip

\noindent
\quad $(\Phi 1)$\quad  
$\Phi$ is strictly convex,

\smallskip

\noindent
\quad $(\Phi 2)$\quad 
$\Phi (f) \geq C f^{1+1/k}$ for  $f \geq 0$ large, with $0 < k < 3/2$,

\smallskip

\noindent
\quad $(\Phi 3)$\quad 
$\Phi (f) \leq C f^{1+1/k'}$ for $f \geq 0$ small, with $0 < k' < 3/2$.

\smallskip

\noindent
For a given 
constant $M>0$ we want to minimize the energy-Casimir functional $\Hc$
over the constraint set
\[ 
\F_M := \left\{ f \in L^1_+(\R^6)\, \mid \,
\int\!\!\!\!\int f dv\,dx = M,\  \ekin(f) + \C(f) < \infty \right\}.
\]
Here  $L^1_+(\R^6)$ denotes the set of non-negative, integrable functions
on $\R^6$. One can show that the potential energy is defined
on this set and the two forms of $\epot$ given in (\ref{epotdef}) are equal,
cf.\  \cite[Lemma 1]{R3}.
Due to conservation of mass the constraint set
$\F_M$ is invariant under solutions of the Vlasov-Poisson system.

The main step in the stability analysis is to
establish the following theorem:
\begin{theorem} \label{exminim}
The energy-Casimir functional $\Hc$ is bounded from below on $\F_M$
with $h_M:=\inf_{\F_M}\Hc <0$.
Let $(f_j) \subset \F_M$ be a minimizing sequence of 
$\Hc$, i.e., $\Hc(f_j) \to h_M$. Then there exists 
a function $f_0\in \F_M$, a subsequence,
again denoted by $(f_j)$ and a sequence $(a_j) \subset \R^3$ of shift vectors
such that 
for the induced gravitational fields
\[
T^{a_j} \nabla  U_{f_j} =\nabla  U_{f_j} (\cdot + a_j) 
\to \nabla U_{f_0} \ \mbox{in}\ L^2 (\R^3),\ j\to \infty.
\]
The state $f_0$ minimizes the energy-Casimir functional: $\Hc(f_0)=h_M$.
\end{theorem}

We shall see shortly that to conclude stability
of the state $f_0$ the theorem is needed in the above form;
mere existence of a minimizer is not sufficient.
A  proof via a reduced functional is given below.
The main difficulty is seen from the following sketch of the
argument:
To obtain a lower bound for the functional on the constraint set is easy,
and by Assumption $(\Phi 2)$ minimizing sequences can be seen to be bounded
in $L^{1+1/k}$. A standard analysis result then implies that such a sequence
has a weakly convergent subsequence, which means that for any test function
$g$ from the dual space $L^{1+k}$ the convergence $\int f_j g \to \int f_0 g$
holds, cf.\  \cite[Sec.~2.18]{LL}. The weak limit $f_0$ is the candidate
for the minimizer, and one has to pass the
limit into the energy-Casimir functional. This is easy for the kinetic energy,
the latter being linear, and for the 
Casimir functional which is convex due to Assumption $(\Phi 1)$
it relies on Mazur's lemma, cf.\  \cite[Sec.~2.13]{LL}. 
The difficult part is the potential energy,
for which one has to prove that the induced gravitational fields
converge strongly in $L^2$.
This problem will we dealt with on the level of the reduced functional
in Section~5.

Since our minimization problem is invariant
under spatial translations we obtain a trivial minimizing sequence 
by shifting a given minimizer in space. 
If for example we shift it off to spatial infinity
we cannot obtain a subsequence which tends weakly to a minimizer,
unless we move with the sequence. Hence the spatial shifts in the
theorem arise from the physical properties of our problem. 

The Euler-Lagrange identity corresponding to our constrained
variational problem implies that any minimizer is a steady
state of the Vlasov-Poisson system. For the proof we refer to 
 \cite{GR3} or  \cite{R5}:

\begin{theorem}\label{minim=ss}
Let $f_0 \in \F_M$ be a minimizer with potential $U_0$. Then
\[
f_0 (x,v)=\left\{
\begin{array}{ccl} 
(\Phi')^{-1}(E_0 - E)&,& E <  E_0,\\
0 &,& \mbox{else} 
\end{array}
\right. \ \  \mbox{where}\ E = \frac{1}{2} |v|^2 + U_0(x)
\]
with Lagrange multiplier $E_0$.
In particular, $f_0$ is a steady state of the Vlasov-Poisson system.
\end{theorem}
For example, the choice (\ref{poly})
yields the polytropic steady state
\[
f_0 (x,v)= \frac{k}{k+1}
\left\{
\begin{array}{ccl} 
(E_0-E)^k&,& E <  E_0,\\
0 &,& \mbox{else}. 
\end{array}
\right.
\]
It should be noted that the assumptions on $\Phi$ easily translate 
into assumptions on the steady state $f_0$ as a function of the 
particle energy, the main one being that this function is strictly 
decreasing on its support.
Various additional properties can be derived for these
minimizers/steady states, in particular they
are necessarily spherically symmetric, 
cf.\  \cite[Thm.~3]{R3} or  \cite[Thm.~2]{R5}. 
Non-symmetric steady states will
be briefly considered in the last section.

To deduce our stability result we expand $\Hc$ about the minimizer
$f_0$:
\be \label{expansion}
\Hc (f)- \Hc (f_0)=d(f,f_0)-\frac{1}{8 \pi}
\int|\nabla U_f - \nabla U_0|^2 dx
\ee
where for $f \in \F_M$,
\beas
d(f,f_0)
&:=& 
\int\!\!\!\!\int \left[\Phi(f)-\Phi(f_0) + E\, (f-f_0)\right]\,dv\,dx\\
&\geq&
\int\!\!\!\!\int \left[ \Phi'(f_0) + (E - E_0)\right]
\,(f-f_0)\,\,dv\,dx \geq 0
\eeas
with $d(f,f_0)=0$ iff $f=f_0$. This is due to the strict convexity of $\Phi$,
and the fact that on the support of $f_0$ the bracket vanishes by 
Theorem~\ref{minim=ss}; note also that
$\int\!\!\!\!\int (f-f_0) = 0$ for $f \in \F_M$. 
For suitable $\Phi$'s, we even have
$d(f,f_0) \geq  c\, \int\!\!\!\!\int |f-f_0|^2$.
The point now is that according to Theorem~\ref{exminim}
the term with the negative
sign in (\ref{expansion}) tends to zero along any
minimizing sequence. This allows us to use the sum of the two
positive definite terms in the expansion as our measure of distance in the 
stability result. As mentioned in the introduction, 
initial data from the space $C^1_c (\R^6)$ of continuously differentiable,
compactly supported functions launch smooth global-in-time solutions
of the Vlasov-Poisson system which preserve all the
physically conserved quantities. As above, $T^a f(x,v):= f(x+a,v)$.
\begin{theorem}\label{stabilitygal}
For any $\epsilon>0$ there exists a $\delta>0$ such that for any 
solution $t \mapsto f(t)$ of the Vlasov-Poisson system 
with $f(0) \in C^1_c (\R^6)\cap \F_M$ the initial estimate
\[
d(f(0),f_0) + \frac{1}{8\pi} \int|\nabla U_{f(0)} - \nabla U_0|^2 dx < \delta
\]
implies that for any $t\geq 0$ there is a shift vector $a\in \R^3$ such that
\[
d(T^a f(t),f_0) + 
\frac{1}{8\pi} \int|T^a \nabla U_{f(t)} - \nabla U_0|^2 dx < \epsilon,
\ t \geq 0,
\]
(provided the minimizer $f_0$ is unique up to shifts).
\end{theorem}
We will comment on the uniqueness assumption (and why it comes in parenthesis)
shortly, but first we give the proof of this result, which is surprisingly 
simple---the difficulty resides in the proof of Theorem~\ref{exminim}.

{\bf Proof:}
Assume the assertion is false. Then there exist 
$\epsilon>0,\ t_j>0,\ f_j(0) \in C^1_c (\R^6)\cap \F_M$ such that for $j\in \N$,
\[
d(f_j(0),f_0) + 
\frac{1}{8\pi} \int |\nabla U_{f_j(0)}-\nabla U_0|^2 dx < \frac{1}{j},
\]
but for any shift vector $a\in \R^3$,
\[
d(T^a f_j(t_j),f_0) +
\frac{1}{8\pi} \int |T^a \nabla U_{f_j(t_j)}-\nabla U_0 |^2 dx 
\geq \epsilon.
\]
Since $\Hc$ is conserved, (\ref{expansion}) and the assumption on 
the initial data implies that
$\Hc (f_j(t_j)) =  \Hc (f_j(0)) \to \Hc (f_0)$,
i.e., $(f_j(t_j)) \subset  \F_M$ is a minimizing sequence.
Hence by Theorem~\ref{exminim},
$\int |\nabla U_{{f_j}(t_j)}-\nabla U_0|^2\to 0$ 
up to subsequences and shifts in $x$, provided that there is
no other minimizer to which this sequence can converge.
By (\ref{expansion}), 
$d(f_j(t_j),f_0)\to 0$ as well, which is the desired
contradiction.
\prfe

Some comments are in order:
For the polytropic steady states one can show that for a given mass
$M$ the minimizer is indeed unique up to shifts, 
as assumed above, cf.\  \cite[Thm.~3]{R3}. 
In general, minimizers do not seem to be unique;
for a numerically verified example of non-uniqueness we refer to
 \cite[Rem.~3 (b)]{R5}. However,  minimizers always seem to be isolated
up to shifts which is sufficient for the above statement to still
hold true, cf.\  \cite[Thm.~3]{R5}. 
If there were a continuum of minimizers then this set of minimizers as
a whole would be stable, cf.\  \cite[p.~242]{GR1}. 
Finally, for a closely related approach to which we will come back in 
the last section it is shown in
 \cite{Sch} that the assertion of Theorem~\ref{stabilitygal} holds
without $f_0$ being unique or isolated. We kept the former assumption 
to make the proof simple.

The spatial shifts appearing in the stability statement are again due
to the spatial invariance of the system: If we perturb
$f_0$ by giving all the particles an additional, fixed velocity,
then in space the corresponding solution travels off from $f_0$ at a 
linear rate in $t$, no matter how small the perturbation was.

A nice feature of the result is that the same quantity is used to measure
the deviation initially and at later times $t$.
In infinite dimensional dynamical systems control in a 
strong norm initially can be necessary
to gain control in a weaker norm at later times.
Whether our concept of distance is appropriate from a physics point of 
view is open to debate---it is simply what comes out of the
energy-Casimir method. For the polytropic steady states
our approach has been extended to yield stability
with respect to the $L^1$-norm of $f$, cf.\  \cite{SS}.
Definitely a weak point is the fact that the
proof is not constructive: Given an $\epsilon$ we do not know how
small the corresponding $\delta$ must be.

\section{The reduced variational problem}
\setcounter{equation}{0}
We wish to factor out the dependence on the velocity variable
in our minimization problem. Starting from a given function $f=f(x,v)$ with
induced spatial density $\rho_f=\rho_f(x)$ we clearly decrease
$\Hc(f)$ by minimizing for each point $x$ over all functions
$g=g(v)$ which have as integral the value $\r_f(x)$. 
This procedure does not affect
the potential energy and reduces the sum of the kinetic energy and the 
Casimir functional into a new functional which no longer depends on $f$ 
directly but only on $\r_f$. More precisely, with
\be \label{phipsi}
\Psi(s) :=
\inf \left\{
\int\left(\frac{1}{2}|v|^2 g + 
\Phi(g)\right)\,dv  \;\mid  \; g \in L^1_+ (\R^3),\ 
\int g(v)\, dv = s \right\}
\ee
we have the estimate $\Hc(f)\geq \Hr(\r_f)$ where 
\[
\Hr(\r) := \int \Psi(\r(x))\, dx + \epot (\r).
\]
We now wish to minimize $\Hr$ over the constraint set
\[
\Rm
:=
\left\{\r \in L^1_+(\R^3) \; \mid \;
\int \Psi (\r) < \infty ,\ 
\int \r = M \right\} .
\]
These constructions owe much to  \cite{Wo}.
Before we analyze the reduced problem we make sure
that we can lift any information gained for the latter back to the
level of the original problem:

\begin{theorem}\label{no-loss} 
For all $f\in \F_M$ the estimate $\Hc(f)\geq \Hr(\r_f)$
holds, with equality if $f=f_0$ is a minimizer. 
Let  $\r_0 \in \Rm$ minimize $\Hr$ and $U_0=U_{\rho_0}$.
Then 
\be \label{redel}
\r_0 = \left\{ 
\begin{array}{ccl}
(\Psi')^{-1}(E_0 - U_0)&,& U_0 < E_0, \\
0 &,& U_0 \geq E_0 
\end{array}
\right.
\ee
with Lagrange multiplier $E_0$, and
\[
f_0
:=
\left\{ 
\begin{array}{ccl}
(\Phi')^{-1}(E_0 - E) &,& E < E_0, \\
0 &,& E \geq E_0. 
\end{array}
\right.\quad E = E(x,v) := \frac{1}{2} |v|^2 + U_0(x),
\]  
lies in $\F_M$ and minimizes $\Hc$.
If on the other hand $f_0 \in \F_M$ minimizes $\Hc$ then
$\rho_{f_0} \in \Rm$ minimizes $\Hr$.
\end{theorem}
Equation (\ref{redel}) is nothing but the Euler-Lagrange
identity for the reduced problem, cf.\  \cite{R5}; 
the theorem is proven in detail in  \cite[Sec.~2]{R4}.

Let us now consider the reduced variational problem in its own right.
The function $\Psi$ defining the reduced functional is 
taken from the following class:\\ 
$\Psi \in C^1 ([0,\infty[)$, $\Psi(0)=0=\Psi'(0)$, and

\smallskip

\noindent
\quad $(\Psi 1)$\quad 
$\Psi$ is strictly convex. 

\smallskip

\noindent
\quad $(\Psi 2)$\quad 
$\Psi (\r) \geq C \r^{1+1/n}$ for $\r \geq 0$ large,  with $0 < n < 3$,

\smallskip

\noindent
\quad $(\Psi 3)$\quad 
$\Psi (\r) \leq C \r^{1+1/n'}$ for $\r \geq 0$ small,  with $0 < n' < 3$.

\smallskip

\noindent
In Section~5 we shall prove the following central result:
\begin{theorem} \label{ccp}
The functional $\Hr$ is bounded from below on $\Rm$.
Let $(\r_j)\subset \Rm$ be a minimizing sequence 
of $\Hr$. Then there exists a sequence of shift vectors 
$(a_j)\subset \R^3$ and a subsequence, again denoted
by $(\r_j)$, such that 
\[
T^{a_j} \r_j \rightharpoonup \r_0
\ \mbox{weakly in}\ L^{1+1/n}(\R^3),\ j \to \infty,
\]
\[
T^{a_j} \nabla U_{\r_j} \to \nabla U_0\ 
\mbox{strongly in}\ L^2(\R^3),\ j \to \infty,
\] 
and $\r_0 \in \Rm$ is a minimizer of $\Hr$.
\end{theorem}
We need to
translate the conditions on $\Psi$
back into conditions on $\Phi$. To do so we denote the Legendre transform 
of a function $h:\R \to ]-\infty,\infty]$ by
\[
\overline{h} (\lambda) := \mathrm{sup}_{r \in \R} (\lambda \, r - h(r)).
\]
If $\Psi$ arises from $\Phi$ by reduction, i.e., by formula (\ref{phipsi}) 
then  
\[
\overline{\Psi} (\lambda)
=
\int \overline{\Phi} \left( \lambda - \frac{1}{2} |v|^2 \right)\, dv.
\] 
This more explicit relation between $\Phi$ and $\Psi$ is established
in  \cite[Sec.~2]{R4}, and it allows us to
show that the assumptions on $\Phi$ imply the ones on $\Psi$
if the parameters $k$ and $n$ are related by $n=k+3/2$, with the same 
relation holding for the primed parameters. 

Theorem~\ref{no-loss} connects our two variational problems
in the appropriate way to derive Theorem~\ref{exminim} 
from Theorem~\ref{ccp}: Firstly, $\Hc$ is bounded from below
on $\F_M$ since this is true for $\Hr$ on $\Rm$.
Let $(f_j)\subset \F_M$ be a minimizing sequence for
$\Hc$. By Theorem~\ref{no-loss}, $(\r_{f_j})\subset \Rm$ is a
minimizing sequence for $\Hr$. 
Again by Theorem~\ref{no-loss} we can lift the
minimizer $\r_0$ of $\Hr$ obtained in Theorem~\ref{ccp}
to a minimizer $f_0$ of $\Hc$, and the proof of Theorem~\ref{exminim}
is complete.

Before we consider some of the ideas involved in the proof of 
Theorem~\ref{ccp} 
we reinterpret it in terms of the Euler-Poisson system.

\section{Pay-off of reduction---The Euler-Poisson system}
\setcounter{equation}{0}
If $\r_0 \in \Rm$ minimizes the reduced functional 
$\Hr$, then $\r_0$ supplemented
with the velocity field $u_0=0$
is a steady state of the Euler-Poisson system 
\[
\dt \rho + \nabla\cdot(\rho u) = 0,
\]
\[
\rho \dt u + \rho (u\cdot \nabla) u = - \nabla p - \rho\, \nabla U,
\]
\[
\lap U = 4 \pi \rho,\ \lim_{|x|\to \infty} U(t,x) = 0,
\]
with equation of state
\[
p=P(\rho) := \rho\, \Psi'(\rho) - \Psi(\rho).
\]
This follows from the Euler-Lagrange identity (\ref{redel}).
Here $u$ and $p$ denote the velocity field and the pressure of an 
ideal, compressible 
fluid with mass density $\r$, and the fluid self-interacts via
its induced gravitational potential $U$. This system is sometimes
used as a simple model for a gaseous, barotropic star.
The beautiful thing now is that obviously
$(\rho_0,u_0=0)$ minimizes the energy
\[
\H (\r,u) :=
\frac{1}{2} \int |u|^2 \rho \, dx + \int \Psi(\r)\, dx + \epot(\r)
\]
of the system, which is a conserved quantity. Expanding as before
we find that
\[
\H (\r,u)- \H (\rho_0,0)
= \frac{1}{2} \int |u|^2 \rho dx + d(\rho,\rho_0)
-\frac{1}{8 \pi}\int |\nabla U_\rho-\nabla U_0 |^2 dx
\]
where for $\rho\in \Rm$,
\[
d(\rho,\rho_0) := \int\left[\Psi(\rho)-\Psi(\rho_0) +
(U_0 - E_0)(\rho-\rho_0)\right] dx \geq 0, 
\]
with equality iff $\r = \r_0$.
The same proof as for the Vlasov-Poisson system implies a stability
result for the Euler-Poisson system---the term with the
unfavorable sign in the expansion again tends to zero along
minimizing sequences, cf.\ Theorem~\ref{ccp}. 
However, there is an important caveat:
While for the Vlasov-Poisson system we have global-in-time solutions
for sufficiently nice data, and these solutions really preserve all the
conserved quantities, no such result is available for the Euler-Poisson system,
and we only obtain a

\noindent
{\bf Conditional stability result:} {\em
For every $\e >0$ there exists a $\delta>0$ such that for every solution
$t \mapsto (\rho(t),u(t))$ with $\rho(0) \in \Rm$ which preserves 
energy and mass
the initial estimate
\[
\frac{1}{2} \int |u(0)|^2 \rho(0)\,dx + d(\rho(0),\rho_0)
 + \frac{1}{8\pi} \int |\nabla U_{\rho(0)}-\nabla U_0 |^2 dx
 < \delta
\]
implies that as long as the solution exists,
\[
\frac{1}{2} \int |u(t)|^2 \rho(t)\,dx + d(\rho(t),\rho_0)
+ \frac{1}{8\pi} \int |\nabla U_{\rho(t)}-\nabla U_0 |^2 dx
 < \epsilon
\]
up to shifts in $x$ (provided the minimizer is unique up to such shifts).}

The same comments as on Theorem~\ref{stabilitygal} apply
also in this case. Because of the above caveat we prefer
not to call this a theorem, although as far as the stability
analysis itself is concerned it is perfectly rigorous.
The open problem is whether a suitable concept of solution
to the initial value problem exists.

Now that minimizers of the reduced functional are identified
as steady states of the Euler-Poisson system it is instructive
to reconsider the reduction procedure leading from the
kinetic to the fluid dynamics picture. First we recall that
for the Legendre transform
\[
h'(\xi) = \eta \Longleftrightarrow 
h(\xi) + \overline{h} (\eta) 
= \xi \eta \Longleftrightarrow \left(\overline{h}\right)'(\eta) = \xi.
\]
If $f_0$ is a minimizer of $\Hc$,
\beas
f_0 
&=& 
(\Phi')^{-1}(E_0 - E) = \left(\overline{\Phi}\right)'(E_0 - E),\\
\rho_0 
&=& 
\int f_0 dv = 
\int \left(\overline{\Phi}\right)'
\left(E_0 - U_0 - \frac{1}{2}|v|^2 \right)\, dv,\\
p_0 
&=& 
\frac{1}{3}\int |v|^2 f_0 dv =
\int \overline{\Phi} \left(E_0 - U_0 - \frac{1}{2}|v|^2\right)\, dv.
\eeas
On the other hand, if $\rho_0$ is a minimizer of $\Hr$,
\beas
\rho_0 
&=& 
(\Psi')^{-1} (E_0 - U_0) = \left(\overline{\Psi}\right)'(E_0 - U_0),\\
p_0 
&=& P(\rho_0) = \rho_0 \Psi' (\rho_0) - \Psi (\rho_0) = 
\overline{\Psi}(\Psi'(\rho_0)) =\overline{\Psi} (E_0 - U_0).
\eeas
In order for these relations on the kinetic and on the fluid level to
fit we have to require that
\[
\overline{\Psi} (\lambda) = \int \overline{\Phi} 
\left(\lambda - \frac{1}{2}|v|^2\right)\, dv
\]
which is the relation between
$\Phi$ and $\Psi$ obtained by our reduction mechanism.

\section{Proof of the existence of minimizers for the reduced problem}
\setcounter{equation}{0}
In this section we present the main arguments leading
to the proof of Theorem~\ref{ccp}. 
Constants denoted by $C$ may only depend on $M$ and $\Psi$
and may change their value from line to line. 
For a set $S\subset \R^3$ the indicator function 
$1_S$ equals 1 on $S$ and vanishes outside. By $B_R$
we denote the ball of radius $R$ about the origin.

\noindent
{\em Step 1: Lower bound for $\Hr$ and weak convergence of minimizing sequences.} 
We need to estimate the negative potential energy against
the positive part of $\Hr$. The Hardy-Littlewood-Sobolev
inequality  \cite[Thm. 4.3]{LL} tells us that
\[
- \epot(\rho) \leq C \|\rho\|_{6/5}^2.
\]
By the restriction on $n$, $1<6/5 < 1+1/n$, and by interpolation
and $(\Psi 2)$,
\[
- \epot(\rho) \leq C 
\nn{\r}_1^{(5-n)/3} \nn{\r}_{1+1/n}^{(n+1)/3}
\leq
C+ C \left(\int \Psi(\r)\, dx\right)^{n/3},\ \r \in \Rm.
\]
Hence on  $\Rm$
\be \label{lowerbound}
\Hr (\r) \geq \int \Psi(\r)\, dx - C -
C \left(\int \Psi(\r)\, dx\right)^{n/3}.
\ee
Since $n<3$ this implies that $\Hr$ is bounded from below on $\Rm$:
\[
h_M := \inf_{\Rm} \Hr > -\infty.
\]
Let $(\r_j)\subset \Rm$ be a minimizing sequence. By (\ref{lowerbound}),
$\int \Psi(\r_j)$ is bounded, and by $(\Psi 2)$
and the fact that $\int \r_j = M$,
the minimizing sequence is bounded in
$L^{1+1/n}(\R^3)$.
Hence we can---after extracting a subsequence---assume that it converges 
weakly to some function $\r_0\in L^{1+1/n}(\R^3)$, i.e., 
for any test function $\sigma\in L^{1+n}(\R^3)$,
$\int \r_j \sigma \to \int \r_0 \sigma$, cf.\  \cite[Sec.~2.18]{LL}. 

As pointed out above, the main difficulty is to prove that 
the induced fields converge strongly in $L^2$; such a result
is referred to as a compactness result. 
It is true if the sequence  $(\r_j)$ remains concentrated:

\noindent
{\em Step 2: Concentration implies compactness.}
Assume that
\be \label{concentration}
\lim_{j \to \infty} \int_{\n{x} \geq R} \r_j = 0,
\ee
for some $R>0$,
i.e., asymptotically the mass remains within the ball $B_{R}$.
We claim that under this assumption 
$\nabla U_{\r_j} \to \nabla U_{\r_0}$ strongly in $L^2$.

By weak convergence, $\r_0 \geq 0$ almost everywhere---if $\r_0$ were 
strictly negative on some set $S$ of positive, finite measure the test 
function $\s=1_S$ 
would yield a contradiction. Moreover, Eqn.~(\ref{concentration})
shows that $\r_0$ vanishes outside the ball $B_{R}$. 
Again by weak convergence, $\int \r_0 =M$. 
The sequence $\s_j := \r_j - \r_0$ converges weakly to $0$ in $L^{1+1/n}$,
$\int |\s_j| \leq M$, and (\ref{concentration}) holds for $|\s_j|$ as well. 
We need to show that
$\nabla U_{\s_j} \to 0$ strongly in $L^2$ 
which is equivalent to 
\be \label{compactness}
I_j := \int\!\!\!\int \frac{\s_j(x) \s_j(y)}{|x-y|} dy\,dx \to 0.
\ee
We choose $R>0$ such that Eqn.\ (\ref{concentration}) applies.
For $\e >0$ we split the domain of integration into the three sets defined by
\beas
&& |x-y| < \e,\\
&& |x-y| \geq \e \ \mbox{and}\ (|x|\geq R \ \mbox{or}\ |y|\geq R),\\
&& |x-y| \geq \e \ \mbox{and}\ |x| < R \ \mbox{and}\ |y| < R,
\eeas
and denote the corresponding contributions to 
$I_j$ by $I_{j,1}, I_{j,2}, I_{j,3}$. 
Since $2n/(n+1) + 2/(n+1) = 2$, Young's inequality  \cite[Thm.~4.2]{LL}
implies that
\[
|I_{j,1}| \leq C ||\s_j||^2_{1+1/n} ||1_{B_\e} |\cdot|^{-1} ||_{(n+1)/2}
\leq C \left(\int_0^\e r^{(3-n)/2}dr\right)^{2/(n+1)} \to 0,
\]
for $\e \to 0$, uniformly in $j$.
Clearly,
\[
|I_{j,2}| \leq \frac{1}{\e} M \int_{|x|> R}|\s_j(x)|\, dx \to 0
\]
as $j\to \infty$, for any fixed $\e>0$.
Finally by H\"older's inequality,
\[ 
|I_{j,3}| = \left|\int \s_j (x) h_j(x)\, dx\right|
\leq ||\s_j||_{1+1/n} ||h_j||_{1+n} \leq C \, ||h_j||_{1+n},
\]
where in a pointwise sense,
\[
h_j(x) 
:=
1_{B_R}(x) \int_{|x-y|\geq \e}1_{B_R}(y) \frac{1}{|x-y|}
\s_j(y)\, dy \to 0
\]
due to the weak convergence of $\s_j$ and the fact that the test function
against which $\s_j$ is integrated here is in $L^{1+n}$.
But since $|h_j| \leq M/\e$ uniformly in $j$
Lebesgue's dominated convergence theorem implies that $h_j \to 0$
in $L^{1+n}$, and the proof of (\ref{compactness}) is complete. 

I wish to thank my student M.~Had\v{z}i\'{c} for the above
rather direct compactness argument  \cite{H}.
 
The next two steps aim to show that minimizing sequences remain
concentrated and do not split into far apart pieces or 
spread out uniformly in space:
 
\noindent 
{\em Step 3: Behavior under rescaling.} 
For $\r \in \Rm$ and $a, b >0$ we define 
$\bar \r(x):= a \r(b x)$.
Then
\[
\int \bar \r\, dx =
a b^{-3} \int \r \, dx,\
\epot(\bar \r) =
a^2 b^{-5} \epot (\r),\
\int \Psi (\bar \r)
=
b^{-3}\int \Psi(a \r)\, dx.
\]

First we fix a bounded
and compactly supported function $\r \in \Rm$ 
and choose $a = b^3$ so that $\bar \r \in \Rm$ as well. 
By $(\Psi 3)$ and since $3/n' > 1$,
\[
\Hr(\bar \r) 
=
b^{-3} \int\Psi(b^3 \r)\, dx + b \, \epot(\r)
\leq
C\, b^{3/n'}   + b \, \epot(\r)
< 0,
\]
for $b$ sufficiently small, and hence $h_M < 0$ for any $M>0$,
i.e., a minimizer---if one exists---is going to be a bound state.

Next we fix two masses
$0<\overline{M} \leq M$. If we take $a=1$ and 
$b=(M/\overline{M})^{1/3} \geq 1$ then for $\r \in \Rm$ and  
$\bar \r \in {\cal R}_{\overline{M}}$ rescaled with these parameters,
\beas 
\Hr (\bar \r)
&=&
b^{-3} \int \Psi(\r)\, dx + b^{-5} \epot (\r) \\ 
&\geq&
b^{-5} \left(\int \Psi(\r)\, dx + \epot (\r) \right) =
\left(\overline{M}/M\right)^{5/3} \Hr (\r).
\eeas
Since for the present choice of $a$ and $b$
the map $\r \mapsto \bar \r$ is one-to-one and onto between
$\Rm$ and ${\cal R}_{\overline{M}}$ 
this estimate gives the following relation between the infima of
our functional for different mass constraints: 
\be \label{mmbar} 
h_{\overline{M}} \geq (\overline{M}/M)^{5/3} h_M,\ 0<\overline{M} \leq M.
\ee

\noindent 
{\em Step 4: Spherically symmetric minimizing sequences remain concentrated.}
In this step we prove Eqn.\ (\ref{concentration}),
but to make matters easier we consider for a moment spherically symmetric
functions $\r \in \Rm$, i.e., $\r(x) = \r(|x|)$. 
For any radius $R>0$ we split $\r$ into the piece supported
in the ball $B_R$ and the rest, i.e.,
\[
\r = \r_1 + \r_2,\ \r_1(x) =0\ \mbox{for}\ |x|>R,\ 
\r_2(x) =0\ \mbox{for}\ |x|\leq R.
\]
Clearly,
\[
\Hr (\r) = \Hr (\r_1) + \Hr (\r_2) - 
\int \frac{\r_1(x)\, \r_2(y)}{\n{x-y}}dx\, dy.
\]
Due to spherical symmetry the potential energy of the 
interaction between the two pieces
can be estimated as
\beas
\int \frac{\r_1(x)\, \r_2(y)}{\n{x-y}}dx\, dy
&=&
- \int U_{\r_1} \r_2  dx\\
&=&
4 \pi \int_R^\infty \frac{4 \pi}{r} \int_0^r \r_1(s)\, s^2 ds\, 
\r_2(r)\,r^2 dr
\leq \frac{(M-m)\, m}{R},
\eeas
where $m=\int \rho_2$ is the mass outside the radius $R$ which we want to make
small along the minimizing sequence. We define 
\[
R_0 := -\frac{3}{5}\frac{M^2}{h_M} > 0
\] 
and use the scaling estimate (\ref{mmbar}) together with the fact that
$h_M<0$ and 
$\xi^{5/3} + (1-\xi)^{5/3} \leq 1-\frac{5}{3} \xi\,(1-\xi)$ 
for $0\leq \xi \leq 1$ to conclude that
\bea \label{splitting}
\Hr (\r)
&\geq&
h_{M-m} + h_{m} - \frac{(M-m)\, m}{R} \nonumber \\
&\geq&
\left[ \left(\frac{M-m}{M}\right)^{5/3} + 
\left(\frac{m}{M}\right)^{5/3} \right]
\, h_M - \frac{(M-m)\, m}{R}\nonumber \\
&\geq&
h_M - \left[\frac{1}{R_0} - \frac{1}{R}\right]\,(M-m)\, m .
\eea
We claim that, if $R>R_0$, then for any
spherically symmetric minimizing sequence $(\r_j) \subset \Rm$ of $\Hr$,
Eqn.\ (\ref{concentration}) holds.
Assume this assertion were false so that up to a subsequence,
\[
\lim_{j \to \infty} \int_{\n{x} \geq R} \r_j = m > 0.
\]
Choose $R_j > R$ such that
\[
m_j:= \int_{\n{x} \geq R_j} \r_j = \frac{1}{2} \int_{\n{x} \geq R} \r_j.
\]
By (\ref{splitting}),
\[
\Hr(\r_j) 
\geq
h_M + \left[\frac{1}{R_0} - \frac{1}{R_j}\right]
\, (M-m_j) \, m_j
\geq
h_M + \left[\frac{1}{R_0} - \frac{1}{R}\right]
\, (M-m_j) \, m_j,
\]
and letting $j \to \infty$ leads to a contradiction.

\noindent 
{\em Step 5: Removing the symmetry assumption.} 
The restriction to spherical
symmetry would mean that stability would only hold against spherically
symmetric perturbations.
Fortunately, the restriction can be removed using a general result due to 
Burchard and Guo. To explain it
we define for a given function $\r\in L^1_+(\R^3)$ 
its spherically symmetric decreasing
rearrangement $\r^\ast$ as the unique spherically symmetric,
decreasing function with the property that for every $\tau\geq 0$
the sup-level-sets 
$\{x \in \R^3 | \r (x) >\tau\}$ and $\{x \in \R^3 | \r^\ast (x) >\tau\}$
have the same volume; the latter set is of course then a ball about the
origin whose radius is determined by the volume of the former. 
The important point
is that for any monotone function $\Psi$ the integral 
$\int \Psi (\r)\, dx$ does not change under such a rearrangement, 
while the potential
energy can only decrease, and it does not decrease if and only if $\r$
is already spherically symmetric (with respect to some center of symmetry) and
decreasing. These facts can be found in  \cite[Ch.~3]{LL}. 
In particular, a minimizer must a posteriori
be spherically symmetric.

Now let $(\r_j)\subset \Rm$ be a 
not necessarily spherically symmetric minimizing sequence.
Obviously, the sequence of spherically symmetric decreasing
rearrangements $(\r_j^\ast)$ is again minimizing.
Hence by the previous steps, up to a subsequence $(\r_j^\ast)$ converges 
weakly to a minimizer $\r_0 = \r_0^\ast$
and 
\[
\nabla U_{\r_j^\ast} \to \nabla U_0 \ \mbox{in} \ L^2,
\ \ \mbox{hence}\ \
\int \Psi (\rho_j^\ast) \to \int \Psi (\rho_0) .
\]
Moreover,
\beas
\epot (\r_j) 
&=&
\Hr (\r_j) - \int\Psi(\r_j)
=
\Hr (\r_j) - \int\Psi(\r_j^\ast)\\
&\to& 
\Hr (\r_0) - \int\Psi(\r_0) = \epot (\r_0).
\eeas
In this situation the result of Burchard and Guo  \cite[Thm.~1]{BG} 
says that up to translations in space
\be \label{fieldconv}
\nabla U_{\r_j} \to \nabla U_0 \ \mbox{in} \ L^2 
\ee

The proof of this general result is by no means easy, and
it is possible to obtain stability against general
perturbations without resorting to it, cf.~ \cite{R3,GR3,R4}.
However, since this general result may be useful for other
problems of this nature we wanted to mention and exploit it here.

\noindent 
{\em Step 6: Proof of Theorem~\ref{ccp}.}
Given a minimizing sequence $(\r_j)$  we already
know that up to a subsequence it converges weakly in $L^{1+1/n}$
to a non-negative limit $\r_0$ of mass $M$.
The functional $\r \mapsto \int \Psi (\r)\, dx$ is convex 
by Assumption $(\Psi 1)$,
so by Mazur's Lemma  \cite[Thm.~2.13]{LL} 
and Fatou's Lemma  \cite[Lemma~1.7]{LL}
\be \label{psiconv}
\int \Psi (\r_0)\, dx \leq 
\limsup_{j\to \infty} \int \Psi (\r_j)\, dx,
\ee
in particular, $\r_0 \in \Rm$.
Together with Eqn.\ (\ref{fieldconv}) this implies that
\[
\Hr (\r_0) \leq \limsup_{i\to \infty} \Hr(\r_i) = h_M
\]
so that $\r_0$ is a minimizer of $\Hr$, and the proof of Theorem~\ref{ccp}
is complete.

\section{Related results, concluding remarks, and open problems}
\setcounter{equation}{0}

In this last section we want to touch upon a number
of questions which may come to mind.

\noindent
{\em The threshold $k=3/2$}.
First we want to ask what happens if we choose the
exponent $k$ in the assumptions $(\Phi 2)$, $(\Phi 3)$
larger than $3/2$. 
This question is answered by the following observations:  
For the Vlasov-Poisson system the Casimir functional $\C$
is preserved. Hence we can pursue an alternative approach, namely to
minimize the energy
\[
\H = \ekin + \epot
\]
under the mass-Casimir constraint
\[
\int\!\!\!\!\int f\, dv\,dx  + \C (f) = M.
\]
As was shown in  \cite{GR3,G2} this works, provided
$\Phi \in C^1 ([0,\infty[)$ with $\Phi(0)=0=\Phi'(0)$
satisfies the assumptions $(\Phi 1)$ (strict convexity) and

\smallskip

\noindent
\quad $(\Phi 2')$\quad 
$\Phi (f) \geq C f^{1+1/k}$ for  $f \geq 0$ large, with $0 < k < 7/2$;

\smallskip

\noindent
$(\Phi 3)$ is not needed any longer.
However, reduction, which combined the kinetic energy and 
the Casimir functional into the new functional $\int \Psi (\r)$,
can no longer work, because
the former two functionals now appear in different places in the 
variational problem. Moreover
for polytropes one can show that
the energy-Casimir functional changes its sign from 
negative to positive at $n=3$ i.e. $k=3/2$. 
If a perturbation leads to an initial datum with positive
energy $\H(f(0)) > 0$ then
\[
\sup \left\{|x| \; \mid \; (x,v) \in \supp f(t)\right\} 
\geq C t,\ t \to \infty.
\]
The analogous result holds for the Euler-Poisson system, except that
for a minimizer the energy in the Euler-Poisson picture equals
the energy-Casimir functional in the kinetic Vlasov-Poisson picture.
Hence the fact 
that $\Hc$ (for the polytropes) changes sign at $k=3/2$ does not
signify any stability change on the kinetic level, but it does 
destroy stability on the fluid level.   

\noindent
{\em Are all relevant isotropic models covered?}
To answer this question it is useful to translate the conditions
on the function $\Phi$ determining the Casimir functional
into conditions on the dependence $f_0(x,v)=\phi(E_0-E)$
of the resulting steady state on the particle energy.
Since $(\Phi')^{-1} = \phi$,
$\Phi$ is strictly convex as required
iff $\phi$ strictly increases; if
$\phi(\eta) \leq C \eta^k$ for $\eta$ large and $\phi(\eta) \geq C \eta^{k'}$
for $\eta>0$ small then $\Phi$ satisfies the growth conditions
$(\Phi 2)$ and $(\Phi 3)$, provided
$0<k,k' <3/2$. In order to satisfy the alternative growth
condition $(\Phi 2')$ we only need to require
$\phi(\eta) \leq C \eta^k$ for $\eta$ large with $0<k <7/2$.
This means that all those polytropic steady states which are
compactly supported and decrease as a function of the local energy
are covered by our results---either by the approach via
reduction with the bonus of the stability of the corresponding
fluid model or, when this is no longer possible, by
the alternative approach mentioned above. The alternative approach
also covers the limiting case $f_0(x,v)=(E_0-E)^{7/2}$, the so-called 
Plummer sphere, cf.~ \cite{GR3}. The minimization property
of the latter is also investigated in  \cite{A}, 
without deducing stability.

An important example from astrophysics which is not yet covered
is King's model where $f_0(x,v)=\exp(E_0-E) - 1$ on its support.
It leads to a Casimir function $\Phi$ growing like $f\ln f$
which is to slow. Possible extensions of the method to include
King's model are currently under investigation.

\noindent
{\em Does the method apply to non-isotropic steady states?}
In the presence of spherical or axial symmetry 
(with respect to the $x_3$-axis) angular momentum quantities like
\[
L:= |x\times v|^2 \ \mbox{or}\ L_3 := x_1 v_2 - x_2 v_1 = (x\times v)_3
\]
are conserved along particle trajectories. If we let the function
$\Phi$ depend also on $L$ or $L_3$ 
the corresponding variational problem has a solution which is a
steady state depending on the additional invariant.
The dependence on $L$ was treated in  \cite{GR1,G1,GR2} while
axially symmetric steady states depending on $L_3$
were treated in  \cite{GR4}. The main assumption again is
the strict monotonicity of the dependence on the local
energy.
There is however one important difference to the isotropic case:
Since a Casimir where $\Phi$ also depends on angular momentum
is preserved by time dependent solutions only if they have
the corresponding symmetry we obtain stability only with respect
to either spherically symmetric or axially symmetric perturbations.
The question of stability of non-isotropic steady states against
non-symmetric perturbations is under investigation.

The method has also been applied to
disk-like steady states, cf.\  \cite{R2}. Here we had to assume that
the perturbations live only in the plane of the disk.
An extension of these results in the spirit of the later developments
reported in the present notes 
or in  \cite{GR3} is under investigation.
Stability against perturbations perpendicular to the
disk is another challenging open problem in this area.

\end{document}